\documentclass[aps,amsmath,amssymb,showkeys]{revtex4}
\usepackage{graphicx}
\usepackage{dcolumn}
\usepackage{bm}

\newcommand{\be}{\begin{equation}}
\newcommand{\ee}{\end{equation}}
\newcommand{\bea}{\begin{eqnarray}}
\newcommand{\eea}{\end{eqnarray}}
\newcommand{\nn}{\nonumber\\}
\def\journal#1#2#3#4{{#1} {\bf #2}, #3 (#4)}

\def\la{\langle}
\def\ra{\rangle}
\def\eq#1{(\ref{#1})}
\def\hf{\frac{1}{2}}
\def\v#1{\mathbf{#1}}

\def\ord#1{{\cal O}\left(#1\right)}

\def\bphi{\bar\phi}

\def\chid{\chi^\dagger}

\begin{document}
\title{Spontaneous breakdown of Lorentz symmetry in scalar QED with higher order derivatives}
\author{Janos Polonyi$^a$, Alicja Siwek$^{ab}$}
\affiliation{$^a$University of Strasbourg,
High Energy Physics Theory Group, CNRS-IPHC,
23 rue du Loess, BP28 67037 Strasbourg Cedex 2, France}
\affiliation{$^b$Wroclaw University of Technology, Institute of Physics, Wybrzeze~Wyspianskiego~27, 50-370 Wroclaw, Poland}
\begin{abstract}
Scalar QED is studied with higher order derivatives for the scalar field kinetic energy.
A local potential is generated for the gauge field due to the covariant derivatives
and the vacuum with non-vanishing expectation value for the scalar field and the 
vector potential is constructed in the leading order saddle point expansion.
This vacuum breaks the global gauge and Lorentz symmetry spontaneously.
The unitarity of time evolution is assured in the physical, positive norm subspace
and the linearized equations of motion are calculated. Goldstone theorem always keeps the 
radiation field massless. A particular model is constructed where the the full set of
standard Maxwell equations is recovered on the tree level thereby relegating the effects
of broken Lorentz symmetry to the level of radiative corrections.
\end{abstract}
\date{\today}
\maketitle

\section{Introduction}
We know no exact equation of motion in physics, all laws are inferred by ignoring
some loosely attached part of the system considered. As a result, the equations
of motion should be tested for the stability of solutions against adding small 
correction terms to the equations. Such an analysis is usually performed in the 
framework of the renormalization group \cite{wilson} and perturbation expansion can be used to establish 
that perturbing terms with higher mass dimension (we use units $c=\hbar=1$) are less 
important at short distances.

Nevertheless, it is an important difference whether the higher mass dimension arises
from field amplitude or space-time derivative because the latter may modify the
tree-level, normal mode structure and generate new degrees of freedom.
In fact, a field theory with a real, single component scalar field
characterized by a Lagrangian containing $n_d$ space-time derivatives of the field contains 
$n_d$ degrees of freedom. Once the new propagating degrees of freedom are present their 
interactions might well be non-negligible due to IR or UV divergences even if the coupling 
constants in the bare Lagrangian are weak. There is another more fundamental change 
generated by these terms, spontaneous symmetry breaking of space-time symmetries
due to an inhomogeneous condensate, the subject of this work. The point is that the renormalization 
group equations usually include quantum fluctuations only. The higher order 
derivatives terms may generate new relevant operators in the IR on the tree-level
which lead to a vacuum with inhomogeneous condensate. We do not embark on a general
renormalization group study here, rather present a simple-minded analysis of the 
symmetry and the quasi-particle content of an Abelian gauge model in the leading order 
saddle point expansion.

If the condensate consists of bosons with non-vanishing momentum, filling up the whole 
quantization volume, then the ``wavy vacuum'' breaks the space-time symmetries, in 
a manner similar to solids where the infinite inertia of the solid prevents the zero modes to 
restore the broken external symmetries. The result, expected from solid state physics, 
is the appearance of several branches of the dispersion relations, different elementary
excitations in the theory. Note that if translation invariance is broken at 
sufficiently short length scale to remain undetectable for the class of observables 
one uses then the vacuum appears homogeneous. We shall see that in models with
gauge symmetry where the covariant derivative is supposed to acquire non-vanishing
value in the condensate the inhomogeneity of the vacuum may be gauged away and we
find a homogeneous condensate which simplifies the model enormously. The result is some 
kind of extension of Higgs-mechanism where the non-vanishing expectation value for 
the gauge field breaks Lorentz symmetry. The resulting Goldstone modes remain in 
the gauge field sector and protect some components of the gauge field against
mass generation. 

The model studied in this work is scalar QED where higher order (covariant) derivative
terms are introduced for the charged scalar field. The higher order terms of this model
can be imagined either as smooth cutoff in defining an UV finite theory or as
originating from the elimination of some heavy particle and approximating
the self energy of a scalar charged particle by a polynomial of finite order in the
momentum. Goldstone theorem protects the electromagnetic field against becoming massive,
Maxwell equations are recovered in the linearized equation of motion,
rendering the Lorentz symmetry breaking effects to radiative corrections.
The rather technical problem of proving unitarity of the model  within the physical,
positive norm subspace is solved within perturbation expansion by assuring real energy 
spectrum for normal modes and preserving the physical subspace, consisting of states
of positive norm during the time evolution.

The dynamical breakdown of space-time symmetries by higher order derivatives has already
been studied in two \cite{afk}, three \cite{afh} and four \cite{afn,afv} dimensional
Euclidean models where periodically modulated condensate has been observed and 
several particle modes have been found corresponding to a single quantum field \cite{afv}.
The present work can be considered as continuation of such inquiries for models defined
in Minkowski space-time and equipped with gauge symmetry. The spontaneous breakdown 
of relativistic symmetries has been considered within the scheme of
emerging photons \cite{bjorken} and the bumblebee models \cite{kostelecky} where an external
Mexican hat potential is assumed for the vector bosons. Our plan is less ambitious and
starts with photons as elementary particles. 

Our results can be best summarized by comparing them with the conventional
Higgs-mechanism where Goldstone mode arising from the spontaneous breakdown of 
global gauge invariance appears in the gauge field which becomes massive.
In our case the relativistic space-time symmetry is broken spontaneously as well,
leaving behind three more Goldstone modes. Two of them are non-vanishing helicity
components of the gauge field and restore the conventional, massless radiation
field of electrodynamics. The third soft mode resides in a certain combination
of the vanishing helicity component of the photon and the scalar field
and is responsible for the preservation of the usual, long range Coulomb
propagator for the temporal component of the gauge field. Therefore, despite
the spontaneous breakdown of internal and external symmetries the free 
propagator and the normal modes of the electromagnetic field are equivalent 
with those of conventional electrodynamics. The symmetry breaking influences 
radiative corrections and the dynamics of the charged scalar field only.

We start in this paper in section \ref{hider} by listing few salient features of scalar 
models with higher order derivatives. The issue of unitarity and the way it 
can be recovered by proving reflecting
positivity in Euclidean space-time is discussed in Section \ref{reflpos}. Our model,
scalar electrodynamics with higher order derivative for the charged scalar field is 
introduced in Section \ref{scaled}. The dynamics is discussed in static temporal gauge where 
the exceptional features of the time component of the gauge field can be dealt with in the easiest manner.
Section \ref{semvac} contains the construction of the vacuum in the leading, tree-level 
order of the saddle point expansion. The stability of the vacuum and the unitarity
within the physical subspace is shown in Section \ref{stabvac}. The particle content
of the theory is defined by the quadratic part of the Lagrangian which is explored in
Section \ref{quphot}. Finally, Section \ref{sum} is reserved to our summary. An Appendix
contains the details of calculating the quadratic part of the action.

\section{Unitarity and higher order derivatives}\label{hider}
Effective theories may or may not be unitary. In fact, the unitarity is lost when 
a particle, retained in an effective theory can lower its energy by the emission 
of other particles which have been eliminated in deriving the effective theory.
Nevertheless, non-unitary effective theory remains a powerful approximation scheme 
when these decay processes are kinematically suppressed and make the life-time
sufficiently long. But one would still prefer to recover the simplicity following
from unitarity in effective theories which tend to be rather complicated.
For instance processes whose energy remains below the mass $M$ of the
particle eliminated should reflect unitary dynamics when considered for
sufficiently long time. Nevertheless the UV divergences and quantum anomalies of the 
underlying theory mix the high energy effects into low energy sector. The most natural 
way of recovering a unitary effective theory is to place the UV cutoff below
the eliminated particle mass, $\Lambda<M$. But this solution is not as simple
as it seems. On one hand, smooth cutoff allows decay processes 
with small but non-vanishing probability, and on the other hand, sharp cutoff leads 
to artificial non-local, acausal dynamics at the length scale $\Lambda^{-1}$ which 
is observable in this case.

The hallmark of effective theories is the appearance of higher order derivatives
in the Lagrangian, reflecting momentum-dependent self energies of quasi-particles 
or form factors. The latter appear in vertices and have mainly perturbative effects. 
But the former are in the quadratic part of the action in the fields and modify the 
structure of quasi-particles. They are sometimes used as a kind
of Pauli-Villars regulator which renders the effective theory UV finite \cite{podolski,villars,leewick}. 
Even though the scale of this smooth cutoff is $M$, the non-unitary processes are not fully
suppressed. Once the effective theory is rendered UV finite we may consider it
as an extension of the class of potentially interesting consistent, microscopic 
models because its UV dynamics is well defined. Motivated by the search of
possible fundamental theories one naturally expects the complete suppression
of non-physical, non-unitary processes. 
 
We consider in this section a model for a neutral scalar particle described by the field 
$\phi(x)$. The interaction vertices will be kept in a momentum independent fashion 
only and the Lagrangian
\be\label{efflagr}
{\cal L}=\hf\phi(x)L(-\Box)\phi(x)-V(\phi^2(x))
\ee
where the real function $L(p^2)$ represents the sum of the kinetic term and a momentum
dependent self energy and is supposed to be a polynomial of order $(p^2)^{n_d}$ which assumes the form
\be\label{lexp}
L(p^2)-V'(\bphi^2)=Z^{-1}\prod_{n=1}^{n_d}(p^2-m^2_n),
\ee
where $Z$ is real, the potential has a minimum at $\phi=\bphi$ and the poles might appear in
complex pairs. The role of the poles $p^2=m^2_n$ can be seen more clearly by means 
of partial fraction decomposition \cite{pais},
\be\label{pvprop}
\frac{Z}{L(p^2)-V'(\bphi^2)}=\sum_n\frac{z_n}{p^2-m_n^2}
\ee
where $z_n=Z/\partial L(m_j^2)/\partial p^2$. We assumed single roots in this equation.
In case a root $p^2=m_n^2$ is of $\ell$-th order then the right hand side may contain
terms $z_n^k/(p^2-m_n^2)^k$ with $1\le k\le\ell$. Complex roots produce complex contributions to
the loop-integrals and lead to exponentially damped or increasing amplitudes in time and 
unitarity can be saved by a graph-by-graph modification of the theory only \cite{cutkosky}.
Another problem is seen for real roots when the kinetic term $L(p^2)$ is a real function and 
it displays slope with alternating sign at its roots. Thus approximately half of the 
contributions to the kinetic energy has the wrong sign, indicating that the Hamiltonian is 
unbounded from below. This instability can be cured by introducing negative norm states 
\cite{leewick} but the unitarity within the physical, positive norm subspace could not be
established in a nonperturbative manner \cite{boulware}. Therefore a careless truncation 
of the self energy may spoil unitarity and stability of the effective theory.

\section{Unitarity and reflection positivity}\label{reflpos}
A proposition to preserve the desired properties was put forward by starting with an 
effective theory in Euclidean space-time \cite{cons}, where it is usually derived 
perturbatively. The effective theory \eq{efflagr} should have a well defined 
Euclidean path integral representation, a condition assured by imposing the constraint $L(-p_E^{02}-\v{p}^2)>0$. 
The safest is to use lattice regularization in Euclidean space-time where higher order 
derivatives can be represented as higher order finite differences.
It is easy to see in lattice regularization that we need new variables to regain the 
usual description for theories with higher order derivatives \cite{hawking}. The 
Kolmogorov-Chapman equation expresses the group structure of the time evolution in the 
Fock-space and can be written as
\be
e^{-S_{t_3-t_1}[\phi^{(3)},\phi^{(1)}]}=\int D[\phi^{(2)}]e^{-S_{t_3-t_2}[\phi^{(3)},\phi^{(2)}]-S_{t_2-t_1}[\phi^{(2)},\phi^{(1)}]}
\ee
where the configurations $\phi^{(j)}$ specify states in the field diagonal representation at
time $t_j$ and the $\exp(-S_{t-t'}[\phi,\phi'])$ denotes the matrix element of the Euclidean
time evolution operator during the time interval $t-t'$. This equation can obviously be derived 
for any theory with nearest neighbor coupling in time. New variables $\phi(x)\to\phi^a(x)$,
$a=1,\ldots,n_d$ which allow us to rewrite the action with higher derivative in a form with
nearest neighbor coupling in time can be introduced in the following manner. Start with a hyper-cubic 
lattice with lattice spacing $a=1$ in each direction and construct an anisotropic lattice where 
the lattice spacing in the time direction is increased to $n_d$ by regrouping $n_d$ time 
slices of the original lattice. A natural choice is $\phi^a(x)=\partial^a_0\phi(x)$, the 
$a$-th order finite difference operator in time acting on the original field where the  
finite difference is calculated from the center of the blocked time slice in a time reversal 
covariant manner assuming odd $n_d$. 
The map $\phi(x)\to\phi^a(x)$ of the Euclidean field variables is an invertible
linear transformation which preserves the lattice regulated action, $S_E[\phi]=S_E[\phi^a]$
and the generator functional,
\bea\label{eqgfo}
Z_E[j]&=&\int D[\phi]e^{-S_E[\phi]+\int dxj\phi}\nn
&=&\prod_a\int D[\phi^a]e^{-S_E[\phi^a]+\sum_xj\phi^0}
\eea
as long as the source is placed at the center of the block time slices.
The transformation preserves its form in Minkowski space-time and provides the mapping whose 
inverse can be used after the Wick rotation of the blocked time slice theory to real time. 

The signature of the norm of states created by the 
operator $\phi^a(x)$ turns out to be $\sigma[\phi^a]=(-1)^a$. In order to preserve the
orthogonality of field eigenvectors, $\la\phi|\phi'\ra=0$ for $\phi(\v{x})\ne\phi'(\v{x})$
we have to use skew-adjoint field operators, which possess
imaginary eigenvalues, in the negative norm sector and $\sigma[\phi]=\pm1$ for self- and skew-adjoint variables.
It is useful to introduce fields with well defined time reversal parity, 
$T\phi(t)=\tau[\phi]\phi(-t)$, giving $\tau[\partial_0^a\phi]=(-1)^a\tau[\phi]$. 
This relation suggests the equivalence of the internal Euclidean time reversal parity and the 
signature of the state created by acting on time reversal invariant vacuum by any time reversal
invariant combination $\psi$ of elementary fields $\phi^a$, 
\be\label{timesign}
\sigma[\psi]=\tau[\psi].
\ee

One has to make sure that unitarity holds within the physical, positive norm subspace, too.
This can be achieved by the reconstruction theorem of axiomatic quantum field theory,
in particular by showing that the main nontrivial condition of the theorem, reflection 
positivity holds in the linear space generated by the action of local operators with 
positive time parity on the vacuum as long as both dynamics and vacuum respect time reversal 
invariance and the boundary conditions $\phi^a(t_f,\v{x})=(-1)^a\phi^a(t_i,\v{x})$ are imposed
where $t_i$ and $t_f$ denote the initial and final time. An important result of the
argument \cite{cons} is the direct verification of Eq. \eq{timesign}.
This relation indicates, as well, that the trajectory of $\phi^a$ in the path integral is 
real or imaginary for $a$ even or odd, respectively. The vacuum may contain condensate 
as long as it is invariant under time reversal. 

This construction gives in the first sight more than expected, it eliminates non-unitarity
altogether for theories \eq{efflagr}-\eq{lexp}. But the tacit assumptions the argument
relies upon are the convergence of the Euclidean path integral and the possibility of its
analytic extension, Wick rotation, back to real time. The former condition (i) imposes $\Re m_n^2>0$. 
The latter assumption requires that the rotation of the frequency contour in the 
loop integrals is carried out without passing singularities in the integrals. 
This conditions excludes poles from the quadrant $\Im m_n^2\cdot\Re m_n^2>0$ of 
the complex energy plane. Since poles come in complex conjugate pairs the remaining complex
poles break time reversal invariance and generate acausality known from the attempts of removing 
self acceleration of point charges in classical electrodynamics \cite{diracr}. Thus time 
reversal invariance restricts the argument to theories where the roots of $L$ are real.

Note that the exclusion of complex poles from the kinetic term restricts the
space-time dependence of the perturbative Green functions to the sum of oscillatory
terms $e^{i\omega t}$ excluding monotonic terms like $e^{\omega t}$. The 
functional space in which the expectation values are constructed is tailored in this
manner and the runaway solutions, characteristic of unstable theories are excluded. 
This is in contrast to classical physics where the integration 
of the equations of motion is performed in an unlimited functional space of trajectories.
Therefore the classical and the quantum, loop-expansion based stability analysis disagree
as far as the time-dependent instabilities are concerned. This eliminates the notorious
instability problem of theories with higher order kinetic term \cite{woodart}.

\section{Scalar electrodynamics}\label{scaled}
An important step towards more realistic models is the extension of previous discussion
for gauge models. We now turn to scalar electrodynamics, defined by the Lagrangian
\be\label{sqedlagr}
{\cal L}=-\frac14\int dxF_{\mu\nu}F^{\mu\nu}+\int dx[\phi^*L(-D^2)\phi-V(\phi^*\phi)],
\ee
with $F_{\mu\nu}=\partial_\mu B_\nu-\partial_\nu B_\mu$ and $D_\mu=\partial_\mu-ieB_\mu$,
$L(z)$ being a polynomial of finite order and is supposed to possess separate time and 
space inversion invariance. In relativistically covariant canonical quantization procedure
one adds a gauge fixing term, ${\cal L}\to{\cal L}-\xi(\partial A)^2/2$, and imposes the canonical 
commutation relations
\be
[A_\mu(t,\v{x}),\Pi_\nu(t,\v{y})]=-ig_{\mu\nu}\delta(\v{x}-\v{y})
\ee
where $\Pi^\mu=\partial{\cal L}/\partial\partial_0 A_\mu$ and $g_{\mu\nu}=(1,-1,-1,-1)$. The 
wrong sign on the right hand side for $\mu=\nu=0$ indicates that temporal photon states 
have negative norm. The Gupta-Bleuler quantization procedure or BRST symmetry can be used 
to prove that usual QED, without higher order derivative is unitary in the physical
subspace, spanned by states with positive norm.

With an $A_0$ field represented by a self-adjoint field operator the field eigenstates are 
not orthogonal. Orthogonality is assured if the operator $A_0$ is skew-adjoint only \cite{cons}. 
The complication, induced by the use of the traditional self-adjoint representation is that 
non-orthogonality renders the path integral expression for the transition amplitudes rather 
complicated. How to recover then the standard path integral representation for gauge theories
in Minkowski space-time? The usual path integral over real field configurations $A_\mu(x)$ can 
easily be found by treating $A_0$ as an auxiliary, non-dynamical field either in static 
temporal or Coulomb gauge. The former will be imposed to establish unitarity in the physical 
subspace because the impact of non-vanishing vacuum expectation value for $A_0$ on the 
dynamics and the similarity with spontaneous symmetry breaking can better be seen in static 
temporal gauge. The latter gauge will be used in clarifying the physical content
of the theory since the dynamical degrees of freedom can be traced easier in Coulomb gauge.

We start with fields defined without initial or final conditions in time for $-\infty<t<\infty$ and
carry out the gauge transformation $A_\mu\to A_\mu+\partial_\mu\alpha$, $\phi\to e^{ie\alpha}\phi$
and $\phi^*\to e^{-ie\alpha}\phi^*$ with
\be\label{tggtr}
\alpha(t,\v{x})=-\int dt'A_0(t',\v{x})
\ee
to arrive at temporal gauge $A_0=0$ where the functional Schr\"odinger representation is constructed, 
using $\v{A}(\v{x})$ as coordinates. The canonical momentum $\v{\Pi}=\partial_0\v{A}=-\v{E}$ satisfies the 
canonical commutation relations $[A_j(\v{x}),\Pi_k(\v{y})]=i\delta_{jk}\delta(\v{x}-\v{y})$.
Gauss' law, $\v{\nabla}\v{E}=\rho$, where $\rho$ is the electric charge density, the equation 
of motion for $A_0$ is lost in this gauge but can be regained as a constraint. In fact,
it can easily be shown by the help of the canonical commutation relations that
\be
G[\alpha]=\int d^3x[\v{\nabla}\alpha(\v{x})\v{E}(\v{x})+\alpha(\v{x})\rho(\v{x})]
\ee
generates static gauge transformations hence it commutes with the gauge invariant 
Hamiltonian $H$, $[G(\v{x}),H]=0$. The average over static gauge transformations,
\be\label{proj}
{\cal P}=\int D[\alpha]e^{i\int d^3x[\v{\nabla}\alpha(\v{x})\v{E}+\alpha(\v{x})\rho(\v{x})]}
\ee
projects into the subspace satisfying Gauss' law for a given static charge distribution $\rho(\v{x})$. 

One is usually interested in transition amplitude between gauge invariant states, the
latter constructed from a gauge-noninvariant representative like a field eigenstate, 
\be
|\v{A},\phi,\phi^*\ra_{sym}={\cal P}|\v{A},\phi,\phi^*\ra.
\ee
It is enough to insert the projection operator $\cal P$ once only in the matrix element,
\bea
\la\v{A}_f,\phi_f,\phi^*_f|e^{-itH}|\v{A}_i,\phi_i,\phi^*_i\ra_{sym}
&=&\la\v{A}_f,\phi_f,\phi^*_f|{\cal P}e^{-itH}|\v{A}_i,\phi_i,\phi^*_i\ra\nn
&=&\int D[\alpha]\la\v{A}_f,\phi_f,\phi^*_f|e^{i\int d^3x[\v{\nabla}\alpha(\v{x})\v{E}+\alpha(\v{x})\rho(\v{x})]}
e^{-itH}|\v{A}_i,\phi_i,\phi^*_i\ra
\eea
and one finds the path integral representation 
\be\label{sttgpi}
\la\v{A}_f,\phi_f,\phi^*_f|e^{-itH}|\v{A}_i,\phi_i,\phi^*_i\ra_{sym}
=\int D[A]D[\phi]D[\phi^*]e^{iS_{st}[A,\phi,\phi^*]}
\ee
where $S_{st}[A,\phi,\phi^*]$ is the usual action in static temporal gauge,
\be\label{sttg}
\partial_0A_0(x)=0,
\ee
and $tA_0(\v{x})=\alpha(\v{x})$ denotes the time-independent integral parameter of 
the projector. The integration is over configurations $\v{A}(t_i,\v{x})=\v{A}_i(\v{x})$,
$\phi(t_i,\v{x})=\phi_i(\v{x})$, $\phi^*(t_i,\v{x})=\phi^*_i(\v{x})$, $\v{A}(t_f,\v{x})=\v{A}_f(\v{x})$,
$\phi(t_f,\v{x})=\phi_f(\v{x})$, $\phi^*(t_f,\v{x})=\phi^*_f(\v{x})$. If the projector is inserted at each time slice of the path integral 
expression for transition amplitude then the gauge invariant action is recovered, 
\be\label{gfreegpi}
\la\v{A}_f,\phi_f,\phi^*_f|e^{-i\Delta tH}{\cal P}\cdots{\cal P}e^{-i\Delta tH}|\v{A}_i,\phi_i,\phi^*_i\ra_{sym}
=\int D[A]D[\phi]D[\phi^*]e^{iS[A,\phi,\phi^*]}
\ee
$\Delta tA_0(t,\v{x})$ playing the role of parameter $\alpha(\v{x})$ in the projector 
inserted at time $t$. 

Temporal gauge, used in the Hamiltonian formalism after Eq. \eq{tggtr} is usually not 
accessible when boundary conditions are imposed in time, as done in path integral expressions. 
Actually the field component $A_0(x)$ represents a true physical variable. We can see this by
noting that $A_0(x)$ cannot be transformed away from the path integral by gauge transformation. 
In fact, setting $A_0=0$ instead of integrating over $A_0(x)$
on the right hand side of Eq. \eq{gfreegpi} removes the projector $\cal P$
on the left hand side and the matrix element is changed, $\la\cdots\ra_{sym}\to\la\cdots\ra$.

A generally applicable gauge choice is static temporal 
gauge, given by Eq. \eq{sttg}. Whatever gauge we use, the Polyakov line
\be\label{abpolyakov}
\Omega(\v{x})=e^{-ie\int_{t_i}^{t_f}dtA_0(t,\v{x})}
\ee
denotes a physical, gauge invariant quantity which prevents us from reaching temporal gauge
as soon as some boundary conditions are 
imposed at the initial and final time. But the integrand of the path integral \eq{sttgpi} remains
unchanged under global gauge transformation of the initial or final state by the phase factor
$1=\exp2\pi i$, represented by the shift
\be\label{abcenter}
A_0(\v{x})\to A_0(\v{x})+\frac{2\pi}{e(t_f-t_i)}.
\ee
Due to this discrete symmetry the integrand in Eq. \eq{gfreegpi} does not depend on the space-time independent
component, $A_0(x)=A_0$ and the variable $A_0$ decouples in the limit $t_f-t_i\to\infty$. 
Nevertheless, the homogeneous component $A_0$ remains a physical parameter when
matrix elements among the vacuum are considered because the vacuum state depends on $A_0$.
In fact, $eA_0$ acts as a chemical potential and one arrives at grand canonical
ensemble where expectation values of observables are saturated by the total charge 
sector of the Fock space which minimizes $H-eA_0\int d^3x\rho$.

\section{Semi-classical vacuum}\label{semvac}
Let us suppose that the model given by Eq. \eq{sqedlagr} is weakly coupled and
saddle point expansion can be used to explore its phase structure. The case of
global symmetry, $e=0$ in the absence of higher order derivative terms
$L(p^2)=p^2$, is well known, the model supports homogeneous condensate
for appropriately chosen local potential. Higher order derivative terms in 
the action may induce a condensation of particles with non-vanishing momentum, 
an inhomogeneous coherent state and a relativistic ``band structure'', 
reminiscent of solid state physics is observed. 
When the interaction with the gauge field is turned on with $L(p^2)=p^2$ then the 
usual Higgs phase can be found. An interesting variant of Higgs mechanism can be
generated by the higher order derivatives terms. The point is that the partial derivatives
are turned into covariant derivatives in the minimal coupling scheme and contain the
connection term which can induce a nontrivial local potential for the gauge field.
The effective interaction, represented by this potential may induce a non-vanishing
expectation value for the gauge field. We call such a vacuum condensate though one 
should keep in mind that it is actually a coherent state only because our gauge 
particle, the photon is neutral and the Bose-Einstein condensation is not possible.

\subsection{Condensate}
We follow the strategy of the saddle point approximation and for this end we 
separate the fields into the sum of saddle point and quantum fluctuations by writing 
$\phi=\bphi+\chi$ and $B_\mu=\bar A_\mu+A_\mu$, the first term in each expression 
representing the saddle point. When a non-vanishing value of the covariant derivative 
\be
-D^2\bphi(x)=k^2\bphi(x)
\ee
is selected for the semi-classical vacuum by the kinetic energy of the charges then a 
gauge transformation can always exchange contributions of the partial 
derivative and the connection term. One possibility is when the eigenvalue $k^2$ in
this equation is provided by the partial derivative alone, $\bphi(x)=\bphi e^{-ipx}$,
$\bar A_\mu=0$. By a suitable gauge transformation we may rearrange the semi-classical 
vacuum into $\bphi(x)=\bphi$, $e\bar A_\mu=k_\mu$. This is a remarkable simplification
offered by gauge invariance, the vacuum consisting of the condensate of particles 
of non-vanishing momentum can be made homogeneous. We exploit this possibility
and assume the homogeneity of the saddle point and the orthogonality of the fluctuations
to the saddle point,
\be\label{orth}
\int dx\chi(x)=\int dxA_\mu(x)=0.
\ee

Note that apart from broken global gauge invariance the gauge field condensate 
leads to the spontaneous breakdown of the Lorentz symmetry. When the function $L(p^2)$
generates spacelike gauge field condensate, $k^2<0$ then Lorentz symmetry is reduced to 
$O(1,2)$ and the excitation spectrum loses rotational invariance. We seek vacuum with 
non-relativistic Galilean $O(3)$ invariance hence we restrict our attention to
models with timelike gauge field condensate,
$e\bar A_\mu=g_{\mu0}k>0$. Hence there will be four combinations 
of fields playing the role of Goldstone bosons when $\bphi,\bar A_\mu\ne0$, 
corresponding to gauge rotations and Lorentz boosts. The number of massless 
particle modes is not necessarily the same. On one hand, it may be smaller because 
either non-relativistic fields have half as many particle modes as their relativistic 
counterparts \cite{leutwyler} or some field combinations may control not 
particle-like excitations, with vanishing residuum in the propagator at the 
``mass-shell''. On the other hand it may be more because higher order derivative 
terms may generate several ``bands''. Global gauge rotation is applied if 
necessary to make the scalar condensate, $\bphi$, real.

\subsection{Fluctuations}
According to section \ref{reflpos} classical stability analysis is sufficient for the 
homogeneous components of the fields and the stability of the 
fluctuations around the vacuum will be verified by checking the spectrum of the elementary 
excitations in quantum theory. The energy-momentum tensor of a theory with polynomial, 
higher order derivative terms can easily be obtained, it is the sum of the usual expression for 
the energy-momentum tensor plus terms containing higher order derivatives of the fields. Therefore the
energy density of the semi-classical homogeneous vacuum characterized by $\bar A_\mu$ and $\bphi$ 
is given by the Lagrangian up to a sign,
\be\label{mina1}
U(e^2\bar A^2,\bphi^2)=-\bphi^2L(e^2\bar A^2)+V(\bphi^2).
\ee
We assume at this point that $L(p^2)$ is bounded from above and it assumes a maximal value at
$p^2=k^2$ thus the minimization with respect to $\bar A^2$,
\be\label{mina2}
0=\frac{\partial U(e^2\bar A^2,\bphi^2)}{\partial e^2\bar A^2}
=-\bphi^2L'(e^2\bar A^2)
\ee
sets $e^2\bar A^2=k^2$ and $e\bar A_\mu=g_{\mu,0}k$ as mentioned above. The separation of 
the kinetic and the potential energy term in the Lagrangian \eq{sqedlagr} for 
the scalar field is not unique, the invariance of the action under the transformation
$L(p^2)\to L(p^2)+\Delta L$, $V(\phi^2)\to V(\phi^2)-\Delta L\phi^2$ can be used to 
set $L(k^2)=0$. We assume the form
\be\label{kinterme}
L(p^2)=-\frac1{k^2}(p^2-k^2)^2,
\ee
the simplest polynomial satisfying our requirements. The scalar condensate $\bphi$ 
is found by minimizing $U(k^2,\phi^2)$, i.e. solving the equation
\be\label{minphi}
0=V'(\bphi^2)-L(k^2)
\ee
with the auxiliary condition that the first non-vanishing derivative of the potential 
at the vacuum is positive.

Once the homogeneous field components are found we turn to the free theory
by considering the quadratic part of the action. We use the decomposition 
$\chi=\chi_1+i\chi_2$, and $\v{A}=\v{n}A_L+\v{A}_T$, $\v{n}=\v{p}/|\v{p}|$, followed
by the separation of the static components $\tilde\chi_a$, $\tilde A_L$ and 
$\tilde{\v{A}}_T$ by writing $\chi_a\to\chi_a+\tilde\chi_a$, $A_L\to A_L+\tilde A_L$ and 
$\v{A}_T\to\v{A}_T+\tilde{\v{A}}_T$. The quadratic action is
written as a sum $S^{(2)}=S^{(2)}+\tilde S^{(2)}$ with
\bea\label{quadraa}
S^{(2)}&=&\hf\int d^4x(\chi_1,\chi_2,A_L,\v{A}_T)\begin{pmatrix}K_{11}&K_{12}&K_{1L}&0\cr
K_{21}&K_{22}&K_{2L}&0\cr K_{L1}&K_{L2}&K_{LL}&0\cr0&0&0&K_{TT}\end{pmatrix}
\begin{pmatrix}\chi_1\cr\chi_2\cr A_L\cr \v{A}_T\end{pmatrix}\nn
\tilde S^{(2)}&=&\frac{t_f-t_i}2\int d^3x(\tilde\chi_1,\tilde\chi_2,\tilde A_0,\tilde A_L,\tilde{\v{A}}_T)
\begin{pmatrix}\tilde K_{11}&0&\tilde K_{10}&\tilde K_{1L}&0\cr
0&\tilde K_{22}&0&\tilde K_{2L}&0\cr 
\tilde K_{01}&0&\tilde K_{00}&0&0\cr 
\tilde K_{L1}&\tilde K_{L2}&0&\tilde K_{LL}&0\cr0&0&0&0&\tilde K_{TT}\end{pmatrix}
\begin{pmatrix}\tilde\chi_1\cr\tilde\chi_2\cr\tilde A_0\cr\tilde A_L\cr\tilde{\v{A}}_T\end{pmatrix}.
\eea
The momentum space representation of the quadratic form,
\be
K(p)=\int dxe^{ip(x-y)}K(x,y)
\ee
is calculated in Appendix \ref{qdramsp} with the result
\bea\label{quadract}
K_{11}&=&L^+_d(p)-4V''\bphi^2=K_{22}\nn
K_{12}&=&iL^-_d(p)=-K_{21}\nn
K_{1L}&=&-|\v{p}|[z(p)L_d(p)]^-=K_{L1}\nn
K_{2L}&=&i|\v{p}|[z(p)L_d(p)]^+=-K_{L2}\nn
K_{LL}&=&\v{p}^2[z^2(p)L_d(p)]^++\omega^2,\nn
K_{TT}&=&\omega^2-\v{p}^2,
\eea
where the notation $f^\pm(p)=f(p)\pm f(-p)$ has been introduced with 
\be\label{shiftedl}
L_d(p)=L((p+e\bar A)^2)-L(k^2),
\ee
$z(p)=e\bphi/(p^2+2\omega k)$
and $p=(\omega,\v{p})$ for the four dimensional fields. The three dimensional,
static sector has quadratic forms $\tilde K(\v{p})=K(p)_{|\omega=0}$, obtained
from Eqs. \eq{quadract} and 
\bea
\tilde K_{10}&=&-\frac{4e\bphi k}{\v{p}^2}L_d(p)_{|\omega=0}=K_{01}\nn
\tilde K_{00}&=&\frac{8e^2\bphi^2k^2}{(\v{p}^2)^2}L_d(p)_{|\omega=0}+\v{p}^2
\eea

\subsection{Unitarity}\label{stabvac}
We turn now to the question of unitarity of the time evolution within the positive norm
subspace of the Fock-space. There are two circumstances requiring to go beyond 
the argument based on the reconstruction theorem for Euclidean theories \cite{osterwalder}.
One is that the manifest $O(4)$/Lorentz invariance of the Euclidean/Minkowski Green functions,
one of the numerous conditions of the theorem is lost in our case. Another point is
that states belonging to excitations generated by the time component of the gauge field 
have negative norm in Minkowski space-time and are thus non-physical. Rather than 
attempting to generalize the reconstruction theorem we choose a simpler
argument, valid in any finite order of the perturbation expansion. 

The partial fraction decomposition of the propagator is now made in terms of $\omega^2$ rather
than $p^2$ and the realness of the one-particle energies guarantees the unitarity 
of the perturbative model within the Fock-space with indefinite norm. 
Perturbation expansion, based on the vacuum with homogeneous fields $\bphi$ and
$\bar A_\mu$ leads to a stable and unitary theory if all solutions of the equation 
$\det K(p)=0$ of the quadratic form $K$ of Eq. \eq{quadraa},
\bea\label{deth}
\det K(p)&=&\frac4{k^4}(\omega^2-\v{p}^2)^2\left\{\omega^{10}-4\omega^8(\v{p}^2+2k^2)
+\omega^6\left[16k^4+16k^2\v{p}^2+6\left(\v{p}^2\right)^2+4V''\bphi^2k^2\right]\right.\nn
&&\left.-\omega^4\left[4V''\bphi^2\left(e^2\v{p}^2\bphi^2+2k^2\v{p}^2-4k^4\right)+4\left(\v{p}^2\right)^3+8k^2\left(\v{p}^2\right)^2\right]\right.\nn
&&\left.+\omega^2\left[64V''^2\bphi^4+4V''\bphi^2k^2\left(\v{p}^2\right)^2+2e^2\bphi^2\left(\v{p}^2\right)^2-4e^2\bphi^2k^2\v{p}^2\right]\right.\nn
&&\left.-2e^2k^2\v{p}^2V''\bphi^3-4e^2\left(\v{p}^2\right)^3V''\bphi^3\right\},
\eea
obtained for the kinetic term \eq{kinterme} have real frequency components, $\omega^2>0$. It 
is easy to see that this expression has negative or complex $\omega^2$ as roots,
there are instable modes in the scalar particle, longitudinal gauge field sector.
These instabilities can be excluded by imposing the condition
\be\label{noge}
V''(\bphi^2)=0
\ee
on the local potential which is not a natural relation, it requires a fine tuning
to cancel the scalar particle scattering amplitude at vanishing momentum. 

According to Goldstone theorem the minimization of the vacuum energy
with respect to the strength of condensate cancels the gap for certain modes.
Goldstone mode arising from the breakdown of global gauge invariance is made by
Eq. \eq{minphi}. As far as the three soft field combinations are concerned, which correspond 
to the breakdown of Lorentz symmetry, let us introduce a mass term for the gauge field by
the extension ${\cal L}\to{\cal L}+m^2B^2/2$ of the Lagrangian \eq{sqedlagr} as in Proca 
theory which leads to the modified potential $U(e^2\bar A^2,\bphi^2)\to U(e^2\bar A^2,\bphi^2)-m^2\bar A^2/2$ 
in Eq. \eq{mina1}. The minimization with respect to the gauge field condensate, 
Eq. \eq{mina2}, generates three soft field combinations. Two of them are the non-vanishing helicity components of transverse 
gauge field even for $m^2\ne0$ and a third is a combination of $\partial_\mu A^\mu$, $\bar A_\mu A^\mu$,
$\chi_1$ and $\chi_2$. To simplify matters we return in our discussion to scalar electrodynamics,
$m^2=0$ where the determinant of Eq. \eq{deth}, whose vanishing identifies the normal mode dispersion relation, reads
\be\label{detst}
\det K(p)=\frac4{k^4}(\omega^2-\v{p}^2)^2\omega^2(\omega^2-2k\omega-\v{p}^2)^2(\omega^2+2k\omega-\v{p}^2)^2.
\ee
The energy spectrum is real, transverse gauge fields make up two Goldstone modes with
$\omega=\pm|\v{p}|$. The scalar field, together with the longitudinal components of the 
gauge field produce the dispersion relations
\be\label{spectrslg}
\omega=\sigma_1k+\sigma_2\sqrt{k^2+\v{p}^2},
\ee
where $\sigma_1,\sigma_2=\pm1$. The choice $\sigma_1\sigma_2=-1$ in Eq. \eq{spectrslg}
belongs to two other Goldstone modes. The determinant \eq{deth} correspond to non-static 
fluctuations, therefore the factor $\omega=0$ in Eq. \eq{detst} is never vanishing.

Once the unitarity has been established in the whole Fock-space let us turn to the physical 
subspace. The argument in Ref. \cite{cons} was presented for Yang-Mills-Higgs model, 
given by the Lagrangian \eq{sqedlagr} though some additional care is required in this case to draw conclusions for 
Minkowski space-time theories. The Wick rotation is more involved for 
gauge than for scalar fields because the norm of state created by $A_0$ changes sign 
during Wick rotation between Euclidean and Minkowski space-time. This leads to the 
following two problems. One has already been mentioned in Section \ref{scaled},
the usual path integration formulas require the orthogonality of the field 
eigenstates and we should use skew-adjoint representation for $A_0$ in Minkowski space-time. 
This amounts to integration over imaginary $A_0$ field which is in an obvious conflict with 
the usual interpretation of $A_0$ as the temporal component of a Hermitian quantum field. 
The solution of this apparent contradiction is well known, the treatment of $A_0$ as a non-dynamical, 
auxiliary variable. This is what happens in static temporal gauge where $A_0$ is the (real)
integral variable of the projection operator to restrict the dynamics into the subspace with 
Gauss' law. Once the real, static $A_0$ configurations are accepted in the path integral
of Eq. \eq{sttgpi} then we may return to the gauge-free case, Eq. \eq{gfreegpi} in the
calculaction of gauge invariant quantities. In other words, in the usual path integral formalism
for real time, available for gauge theories with higher order derivatives, as well, the 
temporal component of the gauge field is better to interpret as an auxilary variable to
handle Gauss' law rather than a quantum field handling physical excitations. 
The situation is reminiscent of conventional QED where elementary excitations, stability, 
renormalizability, etc. are trivial in relativistic gauges but one has to go into another,
physical gauge, usually chosen to be the Coulomb gauge to recover unitarity in the 
physical subspace in an obvious manner.

Another problem, caused by an exceptional feature of $A_0(x)$ during Wick rotation 
is that Eq. \eq{timesign} used to identify the signature of the norm 
is not valid anymore for this component of the gauge field in Minkowski space-time.
A~generalization valid for gauge field is
\be\label{tispsign}
\sigma[\psi]=\tau[\psi]\pi[\psi],
\ee
where $\psi$ is any local combination of the elementary bosonic fields $\partial_0^a\phi$, $\partial_0^a\phi^*$ 
and $\partial_0^a\v{A}$ and space inversion acts as $P\psi(t,\v{x})=\psi(t,-\v{x})$ with 
$\pi[\phi]=-\pi[\v{A}]=1$. PT invariance yields the conservation of $\sigma$ and assures 
unitarity within the positive norm, physical subspace \footnote{The need of space inversion 
invariance was not emphasized in Ref. \cite{cons}.}.

\section{Quasi-photons}\label{quphot}
It has been established so far that our model has unitary time evolution within the positive 
norm subspace and is therefore a physically interpretable. The next question is its physical
content which will be assessed by comparing it with standard electrodynamics.
The usual Higgs-mechanism renders photons massive. The Goldstone modes arising
from the spontaneous breakdown of the Lorentz invariance make three combinations
of the fields soft. Two of them are the transverse, non-vanishing helicity components
of the gauge field and they keep the radiation field massless, just as in standard 
electrodynamics. Two further soft field combinations are made up from the longitudinal
gauge and the scalar field components.

The double pole of \eq{kinterme} may render the normal modes of the scalar field non 
particle-like because scattering amplitude wave packets, constructed by this kind of 
excitations may be vanishing according to the reduction formulas. Thus we take the point of 
view that the scalar field corresponds to so far non-observed excitations and seek the 
dynamics of the gauge field only. To simplify matters further, we ignore radiative
corrections due to the charged scalar field and restrict ourselves to the $\ord{A^2}$ 
part of the action where the normal modes are quasi-photons. We consider below 
two aspects of the model, the number of propagating, dynamical degrees of freedom and
their dispersion relation. It is worthwhile separating two different kinds of
dynamics for the gauge field, first arising through the field strength tensor
in the Maxwell-action, the first term in the Lagrangian \eq{sqedlagr} and second,
coming directly from the connection term of the covariant derivative
in the minimal coupling. The former, field strength tensor dynamics represents 
conventional electrodynamics and the latter, connection term dynamics is the source 
of genuine quantum and topological effects. 

Let us first have a look into the Proca-theory, the simplest model with massive vector field 
and use the standard three-dimensional notation $A^\mu=(\varphi,\v{A})$, $j^\mu=(\rho,\v{j})$, 
$\v{E}=-\v{\nabla}\varphi-\partial_0\v{A}$, $\v{H}=\v{\nabla}\times\v{A}$. We separate the 
transverse and longitudinal components, $\v{A}=\v{A}_T+\v{\nabla}\Phi$, 
$\v{j}=\v{j}_T+\v{\nabla}\kappa$ where current conservation implies 
$\partial_0\rho+\Delta\kappa=0$. The Lagrangian
\be
{\cal L}=\hf\v{E}^2-\hf\v{B}^2-\rho\varphi+\frac{m^2}2(\varphi^2-\v{A}^2)+\v{j}\v{A}
\ee
can be written as $L=L_T+L_{L0}$ where the first and the second term contain
the transverse and longitudinal and temporal components, 
\bea\label{proca}
{\cal L}_T&=&\hf(\partial_0\v{A}_T)^2-\hf\v{B}^2-\frac{m^2}2\v{A}^2_T+\v{j}_T\v{A}_T\nn
{\cal L}_{L0}&=&\hf(\v{\nabla}\varphi+\partial_0\v{\nabla}\Phi)^2+\frac{m^2}2[\varphi^2-(\v{\nabla}\Phi)^2]
-\rho\varphi+\v{\nabla}\Phi\v{\nabla}\kappa.
\eea
As is well known, the temporal component $\varphi$ is not a dynamical degree of freedom 
and can be eliminated by solving its algebraic equation of motion in time,
\be
\varphi=\frac1{m^2-\Delta}(\rho+\partial_0\Delta\Phi),
\ee
without generating non-local effects in time and the resulting effective Lagrangian for $\Phi$ is
\be\label{ltlagr}
{\cal L}_{L0}=-\hf\rho\frac1{m^2-\Delta}\rho+\frac{m^2}2\Phi\frac{\Delta(\Box+m^2)}{m^2-\Delta}\Phi
+\Phi\left[\frac{\Delta}{m^2-\Delta}\partial_0\rho-\Delta\kappa\right].
\ee
For massless photon, $m^2=0$, the equation of motion for $\Phi$ is the current 
conservation and longitudinal photons drop out from the field strength dynamics.
But the mass term, arising from the connection term dynamics may bring the longitudinal 
component back as a genuine dynamical variable. Gauge transformations may make
the separation of auxiliary and truly dynamical variables difficult. For instance,
there are gauges, such as the static temporal gauge, where the 
longitudinal component appears to be dynamical but it drops out from gauge invariant observables.
When higher order derivative terms appear in the connection term dynamics then either 
the temporal or the transverse component of the gauge field may acquire non-trivial dynamics.
The formal gauge invariance always makes the theory redundant therefore one expects 
three dynamical, propagating degrees of freedom for the theory \eq{sqedlagr}
from the photon field, just as in the usual Higgs-mechanism. But their dispersion 
relations differ from those of the Higgs-mechanism, betraying the different 
underlying symmetry breaking patterns.

Let us look into the dispersion relation of the model \eq{sqedlagr} in Coulomb gauge which 
offers a particularly clear view in our model with spontaneously broken Lorentz symmetry. 
The Lagrangian $L=L_T+L_{0m}$ is written as the sum of the transverse part,
given by the first equation in Eqs. \eq{proca}, and the rest whose quadratic part is
\be
{\cal L}_{0m}^{(2)}=\hf(\chi_1,\chi_2,\varphi)K_C
\begin{pmatrix}\chi_1\cr\chi_2\cr\varphi\end{pmatrix}
\ee
where $\chi=\chi_1+i\chi_2$ and
\be
K_C=\begin{pmatrix}L^+_d(p)&iL^-_d(p)&[(p^0+2k)z(p)L_d(p)]^+\cr
-iL^-_d(p)&L^+_d(p)&-i[(p^0+2k)z(p)L_d(p)]^-\cr
[(p^0+2k)z(p)L_d(p)]^+&i[(p^0+2k)z(p)L_d(p)]^-&[(2k+p^0)^2z^2(p)L_d(p)]^++\v{p}^2\end{pmatrix}.
\ee
The dispersion relation is defined by the roots of the determinant
of the quadratic form,
\be\label{detc}
\det K_C(p)=\frac4{k^4}\v{p}^2(\omega^2-2k\omega-\v{p}^2)^2(\omega^2+2k\omega-\v{p}^2)^2.
\ee
Comparing this expression with Eq. \eq{detst}, the determinant of the small fluctuations in 
static temporal gauge apart from the obvious absence of two massless modes, corresponding to non-vanishing 
helicity transverse modes of the gauge field one notices the appearance of a new root, 
$\v{p}^2$, suggesting the emergence of a conventional Coulomb propagator.
One can obtain a more detailed view of the normal modes by the inspection of the
propagators. The inverse of $K_C$ is a full matrix with rather involved matrix elements. 
Matrix elements of $K^{-1}_C$ between the matter field contain the factor 
$(\omega^2-2k\omega-\v{p}^2)^2(\omega^2+2k\omega-\v{p}^2)^2$, indicating the
non-particle like behavior. The matrix elements between the matter field and $\varphi$
have the factors $(\omega^2-2k\omega-\v{p}^2)(\omega^2+2k\omega-\v{p}^2)$ and $\v{p}^2$
in the denominators. Finally, the simplest inverse matrix element is the diagonal 
one for $\varphi$,
\be\label{cprop}
(K^{-1}_C)_{00}=\frac1{\v{p}^2},
\ee
confirming that the factor $\v{p}^2$ in Eq. \eq{detc} corresponds to the unchanged Coulomb law.

The expectation of three dynamical, propagating components for the gauge field, mentioned after
Eq. \eq{ltlagr} above turned out to be wrong and the nontrivial dynamics for the longitudinal component, 
expected by analogy with the Proca case, Eq. \eq{ltlagr} was too naive. The higher order derivative 
terms render the nontrivial dispersion relation for the longitudinal component a gauge artifact
and the usual dispersion relation is recovered for the electromagnetic field.

The surprising simplicity of \eq{cprop} is the result of nontrivial cancellations. This can
easiest be seen by calculating the $A_0$ propagator directly. For this end we eliminate the 
charged field by its equation of motion what is simplest to carry out in the complex $\chi$ basis, where
\be
{\cal L}^{(2)}=\hf(\chi^*,\chi,\varphi)\begin{pmatrix}
K_-&0&K_0\cr0&0&0\cr K_0&0&K_{00}\end{pmatrix}
\begin{pmatrix}\chi\cr\chi^*\cr\varphi\end{pmatrix}
\ee
with $K_-=-2(\Box-2ik\partial_0)^2/k^2$, $K_0=2e\bphi(2k+i\partial_0)(\Box-2ik\partial_0)/k^2$
and $L_{00}=2e^2\bphi^2(\partial_0^2-4k^2)/k^2-\Delta$. The equations of motion for $\chi^*$ and
$\chi$, $0=K_-\chi+K_0A_0$, and $0=\chid K_-+A_0K_0$, used to eliminate the scalar field yield
\be
{\cal L}^{(2)}=\hf\varphi D_{00}^{-1}\varphi
\ee
where
\be
D^{-1}_{00}=K_{00}-\hf(K_0K_-^{-1}K_0+K_0^{tr}K_-^{-1tr}K_0^{tr})
\ee
gives $D^{-1}_{00}=\v{p}^2$ after some cancellations. Therefore the deviation from
usual electrodynamics and the impact of the higher order derivative terms 
are seen by the charged scalar field in our approximation.

\section{Conclusion}\label{sum}
A novel spontaneous symmetry breaking is discussed in the framework of scalar QED
which involves higher order covariant derivatives. One finds non-vanishing expectation value
for the gauge field and unitary, physically acceptable interactions in properly fine 
tuned models.

The unitarity is proven in the physical, positive norm subspace in two steps.
First it is assured in the whole Fock space by fine tuning the self interactions
for the charged scalar field. Second, it is shown that PT invariance makes the
physical subspace closed under time evolution.

The particle content of our model is radically different than the one found in the
conventional Higgs-mechanism. Goldstone theorem renders the radiation field massless.
Furthermore, a particular model is proposed where all components of the gauge field are
massless and Maxwell equations are recovered in the linearized equations of motion. 

We sought in this work a vacuum which supports Galilean invariance, therefore the
temporal component of the gauge field was allowed to develop vacuum expectation value. It
acts as some dynamically generated chemical potential for the charged scalar particle. 
The scalar particle condensate remains electrically neutral due to the equal number 
of particles and anti-particles it contains as a result of the higher order derivative terms
in their dispersion relation.

The status of Lorentz symmetry, broken by the vacuum expectation values to the 
Galilean group, is rather peculiar in the Abelian model. Despite the breakdown of Lorentz invariance Goldstone modes
display relativistic dispersion relations. Furthermore, three components of the gauge field
become Goldstone modes corresponding to the spontenous breakdown of relativistic symmetries 
and hence remain massless even if one starts with massive Proca action for photons. The quadratic part of the 
Lagrangian in the fluctuations of the gauge field is identical to that of QED, leaving the 
Lorentz non-invariant part of the photon dynamics to be generated by radiative corrections.
The deviation of this model from standard electrodynamics is due to radiative corrections
only. 

There are numerous extensions one may consider. Similar models with non-Abelian gauge
symmetry should lead to some massive gauge field components because Goldstone theorem
cannot protect all components of the gauge field anymore against mass generation. 
Using a basis in internal space where the massless gauge bosons are diagonal 
the other, non-commuting components of the gauge field are charged and allow us to 
construct models with an unbroken $U(1)$ subgroup, as in the Standard Model. It remains 
to be seen if natural models, requiring no fine-tuning can be constructed by the eventual
inclusion of charged fermions. Another issue, the scale-dependence of the breakdown
of Lorentz invariance is interesting, too. Being a spontaneous symmetry breaking,
it should be strong at low energy. But some interesting results about non-Lorentz invariant
quadratic terms in gauge theories \cite{nielsen} suggests that certain Lorentz symmetry breaking
parameters of the dynamics tend to be suppressed in the low energy limit. A systematic
renormalization group study of the model would be needed to reveal the true scale 
dependence of this symmetry breaking. Finally, extension
for gravity opens new questions since the spontaneous breakdown of Lorentz symmetry
may generate massive gravitons by a gravitational Higgs-effect.

\appendix
\section{Quadratic action in momentum space}\label{qdramsp}
To find the momentum dependence of the quadratic form $K(p)$ we evaluate the quadratic action 
\eq {quadraa} for the test functions
\bea\label{testchib}
\chi(x)&=&\chi'e^{-ipx},\nn
A_\mu(x)&=&A'_\mu e^{-ipx}
\eea
before gauge fixing for the sake of simplicity. The quadratic form of the 
$\ord{\chi^*\chi}$ part can easily be written as
\be
K_{\chi^*\chi}=2L_d(p)-4V''\bphi^2
\ee
by means of Eq. \eq{minphi} with $L_d(p)$ introduced in Eq. \eq{shiftedl}. 

To find the other terms it is advantageous to represent the higher derivative kinetic term 
of the scalar field as a polynomial,
\be
L(p^2)=\sum_{n=0}^{n_d}c_np^{2n}.
\ee
The block that mixes the scalar and the gauge field originates from the $\ord{B}$ piece in
\be
L(-D^2)\bphi=\sum_{n=0}^{n_d}c_n[-(\partial-ie\bar A-ieA)^2]^n\bphi
\ee
and we find for the $\ord{A\chi}$ contributions
\be
\hf\int dxdy\chi(x)K_{\chi A}(x,y)A(y)=ie\int dx\chi(x)\sum_{n=0}^{n_d}c_n\sum_{\ell=1}^n
(-\bar\Box)\cdots(2A(x)\bar\partial+\partial A(x))\cdots(-\bar\Box)\bphi
\ee
where $\bar\partial_\mu=\partial_0-ie\bar A_\mu$, $\bar\Box=\bar\partial_\mu\bar\partial^\mu$,
and the $\ell$-th factor of the term $\ord{(-D^2)^n}$ is replaced by the $\ord{A}$ part of
$-D^2$ in the right hand side. The choice \eq{testchib} leads to
\be
K_{\chi A}(p)A'=2ie\sum_{n=0}^{n_d}c_n\sum_{\ell=1}^n(p+e\bar A)^2\cdots(-2iA'_0k-ipA')k^2\cdots\bphi,
\ee
written as
\be
K_{\chi A}(p)A'=2e(2A'_0k+pA')k^{-2}\sum_{n=0}^{n_d}c_nk^{2n}\sum_{\ell=0}^{n-1}\left(1+\frac{p^2+2p^0k}{k^2}\right)^\ell\bphi.
\ee
The geometric series can be summed,
\be
K_{\chi A}(p)A'=2e\frac{2A'_0k+pA'}{p^2+2p^0k}\sum_{n=0}^{n_d}c_nk^{2n}\left[(1+\frac{p^2+2p^0k}{k^2})^n-1\right]\bphi,
\ee
and we have
\be
K_{\chi A_\mu}(p)=2e\bphi\left[L_d(p)\frac{2g^{\mu0}k+p^\mu}{p^2+2p^0k}\right]
\ee

The $\ord{A^2}$ quadratic form for real field requires more care. Since it acts on
real field it must be symmetrical. We shall consider a complex plane wave component of the
gauge field in the actual calculation of this term and carry out the symmetrization at the
end only. This term is the sum of two contributions. One of them is the standard Maxwell piece
\be
K^{(1)}_{A_\mu A_\nu}(p)=-T^{\mu\nu}p^2
\ee
where $T^{\mu\nu}=g^{\mu\nu}-p^\mu p^\nu/p^2$ is the projection into the transverse polarization
subspace. The other part is the $\ord{A^2}$ contribution in
\be
\bphi L(-D^2)\bphi=\sum_nc_n\bphi[-(\partial-ie\bar A-ieA)^2]^n\bphi
\ee
which will be written as the sum $K^{(2)}_{AA}(p)+K^{(3)}_{AA}(p)$. The first term stands for 
the $\ord{A^2}$ contributions of the $-D^2$ factor,
\be
A'K^{(2)}_{AA}A'=2e^2\sum_{n=0}^{n_d}c_n\sum_{\ell=1}^n\bphi(-\Box')\cdots A^2(x)\cdots(-\Box')\bphi
\ee
which is vanishing,
\be
K^{(2)}_{A_\mu A_\nu}(p)=2g^{\mu\nu}\bphi^2e^2L'(k^2)=0.
\ee
The other contributions is for the product of two $\ord{A}$ terms,
\bea
A'K^{(3)}_{AA}(p)A'&=&-2e^2\sum_{n=0}^{n_d}c_n\sum_{\ell=1}^{n-1}\sum_{\ell'=\ell+1}^n
\bphi(-\Box')\cdots(2A\partial'+\partial A)\cdots(2A\partial'+\partial A)\cdots(-\Box')\bphi\nn
&=&2e^2\sum_{n=0}^{n_d}c_n\sum_{\ell=1}^{n-1}\sum_{\ell'=\ell+1}^n\bphi k^2\cdots
(2A_0k+pA)\cdots(k^2+p^2+2p^0k)\cdots(2A_0k+pA)\cdots k^2\bphi
\eea
what is written as
\be
A'K^{(3)}_{AA}(p)A'=2e^2\bphi^2(2A_0k+pA)^2k^{-4}\sum_{n=0}^{n_d}c_nk^{2n}\sum_{\ell=1}^{n-1}\sum_{\ell'=\ell+1}^n
\left(1+\frac{p^2+2p^0k}{k^2}\right)^{\ell'-\ell-1}.
\ee
The summation of this geometric series gives
\be
A'K^{(3)}_{AA}(p)A'=2e^2\bphi^2\frac{(2A_0k+pA)^2}{p^2+2p^0k}k^{-2}\sum_{n=0}^{n_d}
c_nk^{2n}\left[\sum_{\ell=1}^{n-1}\left(1+\frac{p^2+2p^0k}{k^2}\right)^{n-\ell}-n+1\right].
\ee
The resulting geometrical series in the square bracket can again be summed with the result
\be
A'K^{(3)}_{AA}(p)A'=2e^2\bphi^2\frac{(2A_0k+pA)^2}{(p^2+2p^0k)^2}\sum_{n=0}^{n_d}
c_nk^{2n}\left[\left(1+\frac{p^2+2p^0k}{k^2}\right)^n-1-n\frac{p^2+2p^0k}{k^2}\right],
\ee
yielding finally
\be
A'K^{(3)}_{AA}(p)A'=2e^2\bphi^2L_d(p)\frac{(2A_0k+pA)^2}{(p^2+2p^0k)^2}
\ee
and
\be
K_{A_\mu A_\nu}(p)=-T^{\mu\nu}p^2+e^2\bphi^2L_d(p)\frac{(2g^{\mu0}k+p^\mu)(2g^{\nu0}k+p^\nu)}{(p^2+2p^0k)^2}
+e^2\bphi^2L_d(-p)\frac{(2g^{\mu0}k-p^\mu)(2g^{\nu0}k-p^\nu)}{(p^2-2p^0k)^2}.
\ee
after symmetrization.


\begin{thebibliography}{99}
\bibitem{wilson} K. G. Wilson, J. Kogut, \journal{Phys. rep.}{12C}{75}{1974}.
\bibitem{afk} P. Azaria, B. Delamotte, T. Jolicoeur, \journal{Phys. Rev. Lett.}{64}{3175}{1990};
M. Dufour Fournier, J. Polonyi, \journal{Phys. Rev.}{D61}{065008}{2000}. 
\bibitem{afh} G. Kohring, R. E. Schrock, \journal{Nucl. Phys.}{B295}{36}{1988};
S. Caracciolo et al, \journal{Nucl. Phys. B Proc. Suppl.}{30}{815}{1993};
J. L. Alonso et al, \journal{Phys. Rev.}{B53}{2537}{1996};
M. L. Plumer, A. Caille, \journal{J. Appl. Phys.}{70}{5961}{1991};
H. Kawamura, \journal{J. Phys. Soc. Jpn.}{61}{1299}{1992};
H. G. Ballesteros et al, \journal{Phys. Lett.}{B378}{207}{1996}; \journal{Nucl. Phys.}{B483}{707}{1997}.
\bibitem{afn} J. L. Alonso et al, \journal{Phys. Lett.}{B376}{148}{1996};
I. Campos, L. A. Fernandez, A. Tarancon, \journal{Phys. Rev.}{D55}{2965}{1997};
H. G. Ballesteros et al, \journal{Phys. Rev.}{D55}{5067}{1997};
Y. Shamir, \journal{Phys. Rev.}{D57}{132}{1998}; J. Fingberg, J. Polonyi, \journal{Nucl. Phys.}{B486}{315}{1997}.
\bibitem{afv} V. Branchina, H. Mohrbach, J. Polonyi, \journal{Phys. Rev.}{D60}{45006}{1999}; \journal{Phys. Rev.}{D60}{45007}{1999}.
\bibitem{bjorken} J. Bjorken, \journal{Ann. Phys.}{24}{174}{1963}; arXiv:hep-th/0111196; arXiv:hep-th/1008.0033.
\bibitem{kostelecky} V. A. Kostelecky, S. Samuel, \journal{Phys. Rev.}{D40}{1886}{1989}.
\bibitem{podolski} B. Podolski, \journal{Phys. Rev.}{62}{68}{1942}.
\bibitem{villars} W. Pauli, F. Villars, \journal{Rev. Mod. Phys.}{32}{434}{1949}.
\bibitem{leewick} T. D. Lee, G. C. Wick, \journal{Nucl. Phys.}{B9}{209}{1969}, \journal{Phys. Rev.}{D2}{1033}{1970}.
\bibitem{pais} A. Pais, G. E. Uhlenbeck, \journal{Phys. Rev.}{79}{145}{1950}.
\bibitem{cutkosky} R. E. Cutkosky, P. V. Ladshoff, D. Olive, J. C. Polinghorne, \journal{Nucl. Phys.}{B12}{281}{1969}.
\bibitem{boulware} D. G. Boulware, D. J. Gross, \journal{Nulc. Phys.}{B233}{1}{1984}.
\bibitem{cons} J. Polonyi, A. Siwek, \journal{Phys. Rev.}{D81}{085040}{2010}.
\bibitem{hawking} S. W. Hawking, T. Hertog, \journal{Phys. Rev.}{D65}{103515}{2002}.
\bibitem{diracr} P. A. M. Dirac, \journal{Proc. Roy. Soc.}{A167}{148}{1938}.
\bibitem{woodart} R. P. Woodard, \journal{Lect. Notes Phys.}{720}{403}{2007}, astro-ph/0601672.
\bibitem{leutwyler} H. Leutwyler, \journal{Phys. Rev.}{D49}{3033}{1994}.
\bibitem{osterwalder} K. Osterwalder, R. Schrader, \journal{Comm. Math. Phys.}{31}{83}{1973};
\journal{ibid.}{42}{281}{1975}.
\bibitem{nielsen} H. B. Nielsen, M. Ninomiya, \journal{Nucl. Phys.}{B141}{153}{1997}.
\end{thebibliography}
\end{document}